\documentclass[final,5p,times,twocolumn]{elsarticle}
\usepackage{color,array,dcolumn}
\usepackage{amsmath}
\usepackage{amssymb}
\usepackage{amsfonts}
\usepackage{graphicx}
\usepackage{slashed}
\usepackage{bbm}
\usepackage{multirow}
\usepackage{color}
\usepackage{slashed}

\newcommand{\eftnopi}{\mbox{EFT($\slashed{\pi}$)}}

\journal{Physics Letters B}

\begin{document}

\begin{frontmatter}

\title{Ground-State Properties of $^{4}$He and $^{16}$O\\
Extrapolated from Lattice QCD with Pionless EFT}

\author[unitn,tifpa]{L. Contessi}
\author[anl]{A. Lovato}
\author[unitn,tifpa]{F. Pederiva}
\author[int]{A. Roggero}
\author[cuny]{J. Kirscher}
\author[ors,ua]{U. van Kolck}
\address[unitn]{Physics Department, University of Trento, Via Sommarive 14, 
I-38123 Trento, Italy}
\address[tifpa]{INFN-TIFPA Trento Institute of Fundamental Physics and 
Applications, Via Sommarive, 14, 38123 Povo TN, Italy}
\address[anl]{Physics Division, Argonne National Laboratory, Argonne, 
Illinois 60439, USA}
\address[int]{Institute for Nuclear Theory, University of Washington, 
Seattle, WA 98195, USA}
\address[cuny]{Department of Physics, The City College of New York, New York, 
NY 10031, USA}
\address[ors]{Institut de Physique Nucl\'eaire, CNRS/IN2P3,
Univ. Paris-Sud, Universit\'e Paris-Saclay, F-91406 Orsay, France}
\address[ua]{Department of Physics, University of Arizona, Tucson, AZ 85721, 
USA}

\begin{abstract}

We extend the prediction range of Pionless Effective Field Theory
with an analysis of the ground state of $^{16}$O in leading order.
To renormalize the theory, we use as input both experimental data 
and lattice QCD predictions of nuclear observables,
which probe the sensitivity of nuclei 
to increased 
quark masses. 
The nuclear many-body Schr\"odinger equation is solved with
the Auxiliary Field Diffusion Monte Carlo method. 
For the first time in a nuclear quantum Monte Carlo
calculation, a linear optimization procedure, which allows 
us to devise an accurate trial wave function with a large number of 
variational parameters, is adopted.
The method yields a binding energy of $^{4}$He
which is in good agreement with experiment at physical pion mass
and with 
lattice calculations at larger pion masses.
At leading order we do not find any evidence of a $^{16}$O state which 
is stable against breakup into four $^4$He, although higher-order terms 
could bind $^{16}$O.
\end{abstract}

\end{frontmatter}

\section{Introduction}
\label{sec:introduction}

Establishing a clear path leading from the fundamental theory of
strong interactions, namely 
Quantum Chromodynamics (QCD), to nuclear observables, 
such as nuclear masses and electroweak transitions, 
is one of the main goals of modern nuclear theory. 
At present, the most reliable numerical technique to perform QCD calculations 
is Lattice QCD (LQCD). It combines recent advances in high-performance 
computing, innovative 
algorithms, and conceptual breakthroughs in nuclear theory to produce 
predictions of nucleon-nucleon scattering, and the binding energies 
and magnetic moments of light nuclei. 
However, there are technical problems,
which have so far limited the applicability of LQCD to 
$A\leq 4$ baryon systems and to artificially large quark masses.
Then, LQCD calculations require significantly smaller computational 
resources to yield meaningful signal-to-noise ratios.
In this paper, we consider as examples LQCD data sets comprised of
binding energies obtained at pion masses of 
$m_\pi \simeq 805$ MeV \cite{Beane:2012vq} and 
$m_\pi \simeq 510$ MeV \cite{Yamazaki:2012hi}.

The link between 
QCD and the entire nuclear landscape
is a potential whose systematic derivation was developed in the framework of
effective field theory (EFT) in the last two 
decades~\cite{Bedaque:2002mn,Epelbaum:2008ga,Machleidt:2011zz}. 
This is achieved by exploiting a separation 
between ``hard''~($M$) and ``soft''~($Q$) momentum scales.
The active degrees of freedom at soft scales are 
hadrons whose interactions are consistent with QCD. 
Effective potentials and currents 
are derived from the most general 
Lagrangian constrained by the QCD symmetries,
and employed with standard few- and many-body techniques to make predictions 
for nuclear observables in a systematic expansion in $Q/M$.
The interactions strengths carry information about the details of
the QCD dynamics, and can be obtained by matching
observables calculated in EFT and LQCD.

The aim of this work is the first extension of this program to the realm of 
medium-heavy nuclei. By using Pionless EFT (\eftnopi)
coupled to the Auxiliary Field Diffusion Monte Carlo (AFDMC) method 
\cite{Schmidt:1999lik} we analyze the connection between the ground state of $^{16}$O and
its nucleon constituents. Beside physical data, the consideration
of higher quark-mass input allows us to investigate the 
sensitivity of $^{16}$O stability to the pion mass.
The usefulness of \eftnopi~\cite{Bedaque:2002mn,Epelbaum:2008ga}
for the analysis of LQCD calculations has been discussed 
previously~\cite{Barnea:2013uqa,Beane:2015yha,Kirscher:2015yda}.

Whether \eftnopi~can be extended to real and lattice nuclei in the 
medium-mass region is an open question. 
For physical pion mass, convergence has been demonstrated in leading orders
for the low-energy properties of $A=2,3$ systems 
\cite{Chen:1999tn,Kong:1999sf,Vanasse:2013sda,Konig:2015aka}.
Counterintuitively, the binding energy of the $A=4$ ground state was
found in good agreement with experiment at leading order 
(LO)~\cite{Platter:2004zs},
and even the $A=6$ ground state comes out reasonably well at this
order~\cite{Stetcu:2006ey}.

A similar binding energy per nucleon for $^4$He ($\simeq 7$ MeV) and 
$^{16}$O ($\simeq 8$ MeV) suggests that \eftnopi~might 
converge for heavier systems.
However, the difference in total binding energy 
between the two systems is quite large. Moreover, many-body effects become 
stronger, and quantum correlations might substantially change the picture.
We chose $^{16}$O for mainly two reasons:
First, because it is a doubly magic nucleus,
thereby reducing the technical difficulties related to 
the construction of wave functions with the correct quantum numbers and 
symmetries. Second, its central density is sufficiently high to probe saturation
properties and thereby serve as a model for even heavier nuclei.

In practice, this first calculation of an $A>6$ system
in \eftnopi~at LO is carried out as follows. 
In order to show renormalizability, we use potentials characterized by cutoffs
up to $\simeq 1.5$~GeV.
This introduces non-trivial difficulties in solving the Schr\"odinger equation 
due to the rapidly changing behavior of the wave functions. To this aim, 
we developed an efficient linear optimization scheme to devise high-quality 
variational wave functions. Those have been employed as a starting point for 
the imaginary-time projection of AFDMC which 
enhances the ground-state component of the trial wave function.  Finally, to 
alleviate the sign problem, we have also performed unconstrained propagations 
and studied their convergence pattern.
We show that, thanks to these developments,
the errors from the AFDMC calculation 
are now much smaller than the uncertainty originating from
the \eftnopi~truncation and the LQCD input.
The door is open for higher-order calculations with
future, more precise LQCD input.

The rest of the paper is organized as follows:
in Sec. \ref{sec:piless}
we will briefly review the properties of \eftnopi~that are relevant 
for our discussion;
in Sec. \ref{sec:MCmethod}
the methodological aspect of the calculations will be discussed;
in Sec. \ref{sec:results}
we will present and discuss our results;
and finally 
Sec. \ref{sec:conclusions}
is devoted to conclusions.
An appendix describes the way we estimate errors.

\section{Pionless Effective Field Theory}
\label{sec:piless}

An EFT is a reformulation of an underlying theory 
in terms of degrees of freedom relevant to the problem at hand,
which interact via operators that obey symmetries compatible with the 
original interactions.
These operators are part of a controlled expansion in a suitable small
parameter which encapsulates a separation of scales in the system.

The 
relativistic, underlying theory, which 
presumably allows for the description of nuclei from first principles, is QCD. 
Low-energy processes in nuclear physics involve small enough momenta to 
justify the use of a nonrelativistic approach. Consequently, 
nucleon number is conserved and 
nuclear dynamics can be described within nonrelativistic many-body theory, 
while the 
strong nuclear potential needs to include only parity and time-reversal 
conserving operators. All relativistic 
corrections are sub-leading. 

In this paper we are interested in the ground states of nuclei.
The characteristic momentum in a two-body bound state of binding energy
$B_2$ is given by the location
of the pole of the $S$ matrix in the complex momentum plane,
$Q_2=\sqrt{m_N B_2}$, where $m_N$ is the nucleon mass.
To our knowledge, there is no consensus for a definition of an analogous
characteristic momentum for larger nuclei bound by $B_A$;
as an estimate one can use a generalization where each nucleon contributes
equally,
\begin{equation}
Q_A=\sqrt{2 m_N\frac{B_A}{A}}\, .
\label{QA}
\end{equation}
For lattice $^4$He at $m_\pi=805$ MeV ($m_\pi=510$ MeV), where
$B_4\simeq 110$ MeV \cite{Beane:2012vq} 
($B_4\simeq 40$ MeV \cite{Yamazaki:2012hi}) with
$m_N\simeq 1600$ MeV ($m_N\simeq 1300$ MeV), 
this estimate gives $Q_4/m_\pi \simeq 0.4 \; (0.3)$. 
Thus, the typical momentum is small
in comparison not only with $m_N$, but also with $m_\pi$,
allowing for a description where pion exchanges are treated as unresolved
contact interactions. The case is less clear-cut in the
real world, where $Q_4/m_\pi \simeq 0.8$, but the results
at LO~\cite{Platter:2004zs} suggest that Eq. \eqref{QA} 
overestimates the typical momentum.
In fact, a similar inference can be made from results of the analog
of \eftnopi~for $^4$He atomic clusters \cite{Bazak:2016wxm}.
At physical $m_\pi$, the binding energy per particle in $^{16}$O is 
similar to that in $^4$He, so \eftnopi~might converge for this nucleus as well.

With pions integrated out, $m_\pi$ gives an upper bound
on the breakdown scale $M$ of the EFT.
The momentum associated with nucleon excitations of mass $m_R$ is given by
Eq. \eqref{QA} with $B_A/A\to m_R-m_N$. For the lightest excitation,
the Delta isobar, $m_\Delta-m_N$ decreases as $m_\pi$ increases. 
For $m_\pi=805$ MeV, $\sqrt{2m_N(m_\Delta-m_N)}$ becomes comparable
to $m_\pi$. Thus, throughout the considered range of pion masses
nucleon resonances can also be treated as short-range effects,
and nucleons are indeed the only relevant degrees of freedom.

At leading order in $1/M$, the \eftnopi~Lagrangian 
\cite{Bedaque:1997qi,Chen:1999tn,Bedaque:1999ve}
consists of the nucleon kinetic term, two two-nucleon contact interactions,
and one three-nucleon contact interaction.
The singularity of these interactions leads to 
divergences that need to be dealt with by
regularization and renormalization.
Here, as in Refs. \cite{Barnea:2013uqa,Kirscher:2015yda},
we use a Gaussian regulator that
suppresses transferred momenta above an ultraviolet cutoff $\Lambda$.
This choice ensures that the Lagrangian can be transformed into a Hamiltonian 
containing only local potentials, suitable to be used within AFDMC. 
The Hamiltonian in coordinate space 
reads~\cite{Barnea:2013uqa,Kirscher:2015yda}
\begin{eqnarray}
H_{LO} &=&-\sum_i \frac{{\vec{\nabla}_i^2}}{2m_N}
+\sum_{i<j}
{\left(C_1  + C_2\, \vec{\sigma_i}\cdot\vec{\sigma_j}\right) 
e^{-r_{ij}^2\Lambda^2 / 4 }}
\nonumber\\
&&+D_0 \sum_{i<j<k} \sum_{\text{cyc}} 
{e^{-\left(r_{ik}^2+r_{ij}^2\right)\Lambda^2/4}}\,,
\label{ham}
\end{eqnarray}
where the sums are over, respectively, nucleons, nucleon pairs,
and nucleon triplets, and $\sum_{\text{cyc}}$ stands for the cyclic permutation of
$i$, $j$, and $k$.
Dependence on the arbitrary regulator choice
is eliminated by allowing the interaction strengths,
or low-energy constants (LECs),
$C_{1}(\Lambda)$, $C_{2}(\Lambda)$ and $D_0(\Lambda)$ to depend on $\Lambda$.

To solve the two-nucleon system, in principle one iterates 
interactions only in the channels containing $S$-matrix poles
within the convergence range of the theory \cite{vanKolck:1998}.
Since two nucleons have a bound state in the $^3S_1$ channel and 
a shallow virtual state (which becomes a bound state as $m_\pi$ 
increases \cite{Beane:2012vq,Yamazaki:2012hi}) 
in the $^1S_0$ channel, one needs to include two interactions 
at LO and treat them non-perturbatively.
In Eq. \eqref{ham} we chose the operator basis $1$ and 
$\vec{\sigma_i}\cdot\vec{\sigma_j}$, but it 
can by replaced by any other form equivalent under
Fierz transformations in SU(2). 
All these possible choices are equally convenient for 
an AFDMC calculation.

When the three-body problem is solved with these interactions,
renormalizability requires a contact three-nucleon force at LO~\cite{Bedaque:1998kg}.
As for the two-body interactions, there is some freedom in choosing the 
operator to include in the Hamiltonian formulation of the three-body force. 
For simplicity we use a central potential,
which makes obvious the Wigner spin-isospin symmetry (SU(4)) of this force.

For renormalization at LO, we ensure that three uncorrelated observables 
are $\Lambda$-independent. 
Here we follow Ref. \cite{Kirscher:2015yda} 
and choose these observables as
the deuteron and triton binding energies and, 
for physical (unphysical) pion mass(es) the $^1S_0$ scattering length
(dineutron binding energy).
The LECs' dependence on the cutoff can be found in 
Ref. \cite{Kirscher:2015yda}.
In particular, because $C_1(\Lambda)\gg C_2(\Lambda)$,
the LO Hamiltonian has an approximate SU(4) symmetry.

Interactions with more derivatives represent higher orders.
For example, at NLO the first two-body energy corrections appear
in the form of two-derivative contact interactions \cite{vanKolck:1998}.
For ground states electromagnetic interactions are also sub-leading,
starting at NLO with the Coulomb interaction \cite{Konig:2015aka}.
Since sub-leading interactions are suppressed by powers of $M$,
they should be included as perturbations.
Treating them non-perturbatively, like the LO terms, is problematic 
as the iteration of sub-leading terms usually destroys renormalizability.
NLO interactions have been dealt with fully perturbatively for 
$A=2$ \cite{Chen:1999tn,Kong:1999sf}
and $A=3$ \cite{Vanasse:2013sda,Konig:2015aka}. 
So far, no perturbative NLO calculation has been conducted
in \eftnopi~for $A\ge 4$. We therefore limit ourselves to LO in this
first foray into medium-mass nuclei.

One feature of the EFT approach is a better understanding of the systematic
uncertainties which reduce the accuracy of predicted nuclear observables.
Apart from the errors germane to AFDMC,
the EFT at LO is expected to be affected by
systematic, relative errors of $\mathcal{O}(Q_A/M, Q_A/\Lambda)$
plus ``measurement'' uncertainties in the LECs.

For observables that were not used as input, regularization
introduces an error proportional to the inverse of the cutoff.
For example, different Fierz-reordered forms of the potential
only give the same results for large cutoffs. In order to minimize 
the regularization error, we fit finite-cutoff results with
\begin{equation}
O_\Lambda = O  + \frac{\mathcal{C}_0}{\Lambda}
+ \frac{\mathcal{C}_1}{\Lambda^2} + \cdots
\, ,
\label{eq:cut_off_extrapolation}
\end{equation}
where $O$ is the observable at $\Lambda \rightarrow \infty$, 
while the parameters $\mathcal{C}_0$, 
$\mathcal{C}_1$, $\dots$, are specific for each observable.
The number of powers of $\Lambda$ needed to perform a meaningful extrapolation 
is not 
known {\it a priori}. The standard prescription consists in truncating the 
expansion when adding additional powers of $1/\Lambda$ no longer influences 
$O$.
In a renormalizable theory, observables converge in the $\Lambda \to \infty$
limit to a value that must not be confused with a precise physical result.
Observables are unavoidably plagued by the truncation error,
which cannot be reduced without a next-order calculation.

The truncation of a natural EFT expansion at order $n$ allows
for a residual error proportional to $(Q/M)^{n+1}$, where the constant of 
proportionality depends on the specific observable. 
Truncation errors are more difficult to assess here because 
the scales $M$ and $Q_A$ are not well known.
Assuming $M\sim m_\pi$ and $Q_A\sim Q_{3,4}$, as given in Eq.~\eqref{QA}, 
one estimates $Q_A/M \sim 1/3$ for physical pions. 
An alternative that does not rely on an estimate for $Q_A$
uses cutoff variation to place a lower bound
on the truncation error. 
The residual cutoff dependence cannot be
distinguished from higher-order contributions.
Assuming that for the observable of interest the leading missing
power of $1/M$ is the same as the leading power of $1/\Lambda$, varying $\Lambda$ from
$M$ to much larger values gives an estimate of the truncation error.
For another technique to estimate the EFT truncation error, see for example
Ref. \cite{Binder:2015mbz}. 

Finally, experimental and numerical LQCD uncertainties are transcribed 
through the renormalization of the LECs. 
While this is not an important issue for the physical data,
LQCD ``measurements'' carry a significant uncertainty which could 
dramatically affect EFT predictions.
Estimating their effects would require a huge computational effort,
as the calculation would have to be repeated for various combinations
of the extreme values the LECs can take. 
Since the pertinent errors \cite{Beane:2012vq,Yamazaki:2012hi}
are comparable to the LO truncation error, this effort is not yet justified.
We will limit ourselves to show that the Monte Carlo errors discussed
in the next section have reached a point where they are not an obstacle
to future higher-order calculations.
At that point,
a more detailed analysis of the propagation 
of ``measurement'' errors at unphysical pion masses will be required.

\section{Monte Carlo Method}
\label{sec:MCmethod}

Quantum Monte Carlo (QMC) 
methods allow for solving the time-independent Schr\"odinger equation of 
a many-body system, providing an accurate estimate of the statistical error 
of the calculation. For light nuclei, QMC and, in particular, 
Green's Function Monte Carlo (GFMC) methods have been successfully exploited 
to carry out calculations of nuclear properties, based on realistic 
Hamiltonians including two- and three-nucleon potentials, and consistent 
one- and two-body meson-exchange currents \cite{Carlson:2014vla}.

Because the GFMC method involves a sum over spin and isospin, its computational
requirements grow exponentially with the number of particles. 
Over the past two decades AFDMC \cite{Schmidt:1999lik} has emerged as 
a more efficient algorithm for dealing with larger nuclear 
systems \cite{Gandolfi:2014ewa}, but only for somewhat simplified interactions.
Within AFDMC, the spin-isospin degrees of freedom are described by 
single-particle spinors, the amplitudes of which are sampled using 
Monte Carlo techniques, and the coordinate-space diffusion in GFMC is extended
 to include diffusion in spin and isospin spaces. 
Both GFMC and AFDMC have no difficulties in using realistic two- and three-body
forces; the interactions are not required to be soft and hence can generate 
wave functions with high-momentum components. 
This is particularly relevant to analyze the cutoff dependence of 
observables, as relatively large values for $\Lambda$ are to be considered
in order to confirm renormalizability. 

QMC methods employ an imaginary-time 
($\tau$)
propagation in order to extract the lowest many-body state $\Psi_0$ from a 
given initial trial wave function $\Psi_T$:
\begin{equation}
|\Psi_0\rangle=\lim_{\tau\to\infty} e^{-(H-E_T)\tau} | \Psi_T\rangle\, .
\label{eq:tau_ev}
\end{equation}
In the above equation $E_T$ is a parameter that controls the normalization of 
the wave function and $H$ is the Hamiltonian of the system. In order to 
efficiently deal with spin-isospin dependent Hamiltonians, the 
Hubbard-Stratonovich transformation is applied to the quadratic spin and 
isospin operators entering the 
imaginary-time propagator to make them linear. As a consequence, the 
computational cost of the calculation is reduced from exponential to 
polynomial in the number of particles, allowing for the study of 
many-nucleon systems.

The standard form of the wave function used in QMC calculations of light nuclei
reads
\begin{equation}
\langle X |\Psi_T\rangle = 
\langle X |\Big(\prod_{i<j<k}U_{ijk}\Big)\Big(\prod_{i<j}F_{ij}\Big)|\Phi\rangle\, ,
\label{eq:trial}
\end{equation}
where $X=\{x_1\dots x_A\}$ and the generalized coordinate 
$x_i= \{r_i,\sigma_i,\tau_i\}$ represents 
the position, spin, and isospin
variables of the $i$-th nucleon. 
The long-range behavior of the wave function is described by  
the Slater determinant
\begin{equation}
\langle X |\Phi\rangle=\mathcal{A}\{\phi_{\alpha_1}(x_1),\dots,\phi_{\alpha_A}(x_A)\}
\, .
\end{equation}
The symbol $\mathcal{A}$ denotes the antisymmetrization operator and 
$\alpha$ denotes the quantum numbers of the 
single-particle orbitals, given by
\begin{equation}
\phi_{\alpha}(x)=R_{nl}(r) \, Y_{\ell \ell_z}(\hat{r})\, 
\, \chi_{ss_z}(\sigma)\, \chi_{\tau \tau_z}(\tau)
\, ,
\end{equation}
where $R_{nl}(r)$ is the radial function, $Y_{\ell \ell_z}(\hat{r})$ is the 
spherical harmonic, and $\chi_{ss_z}(\sigma)$ and $\chi_{\tau \tau_z}(\tau)$ are 
the complex spinors describing the 
spin and isospin of the single-particle state. 

In both the GFMC and the latest AFDMC calculations spin-isospin dependent 
correlations 
$F_{ij}$ and $U_{ijk}$ are 
usually adopted. However, these are not necessary for 
this work. In fact, the two-body LO 
pionless nuclear potential considered in this work does not contain tensor 
or spin-orbit operators. 
In addition, the LEC proportional to the spin-dependent component of the 
interaction is much smaller than the one of the central channel,
$C_2 \ll C_1$. 
Finally, 
Fierz transformations allow us to consider the purely central 
three-body force 
in Eq. \eqref{ham}. As a consequence, we can limit ourselves to 
spin-isospin independent two- and three-body correlations only,
\begin{align}
F_{ij} = f(r_{ij}) \ , \  U_{ijk}=1-\sum_{\text{cyc}} u(r_{ij})u(r_{ik})u(r_{jk}) \, ,
\end{align}
where $f(r)$ and $u(r)$ are functions of the radius only.

The radial functions of the orbitals as well as those entering the 
two-
and 
three-body Jastrow correlations are determined minimizing the 
ground-state expectation value of the 
Hamiltonian,
\begin{equation}
E_V=\frac{\langle \Psi_T | H | \Psi_T \rangle}{\langle \Psi_T | \Psi_T \rangle}
\, .
\end{equation}
In standard nuclear Variational Monte Carlo (VMC) and GFMC calculations 
the minimization is usually done adopting 
a ``hand-waving'' procedure, while in more recent AFDMC calculations 
the stochastic reconfiguration (SR) method~\cite{Sorella:2005} 
has been adopted. In both cases the number of variational parameters is reduced
by first minimizing the two-body cluster contribution to the energy per 
particle, as described in Refs.~\cite{Lagaris:1981mn,Arriaga:1995ue}. 
In this work we adopt, for the first time in a nuclear QMC calculation, 
the more advanced {\it linear method}
(LM)~\cite{Toulouse:2007}, which allows us to deal with a much larger number 
of variational parameters. 

Within the LM, at each optimization step we expand the normalized trial 
wave function 
\begin{equation}
| \bar{\Psi}_T (\mathbf{p}) \rangle = \frac{| \Psi_T (\mathbf{p}) \rangle}
{\sqrt{\langle \Psi_T (\mathbf{p}) | \Psi_T (\mathbf{p}) \rangle }}
\end{equation}
at first order around the current set of variational parameters 
$\mathbf{p}^0=\{p_1^0,\dots,p_{N_p}^0\}$,
\begin{equation}
\label{eq:lm_expansion}
 | \bar{\Psi}_T^\text{lin}(\mathbf{p} ) \rangle = 
 | \bar{\Psi}_T (\mathbf{p}^0) \rangle 
 + \sum_{i=1}^{N_p} \Delta p_i |\bar{\Psi}_T^i  (\mathbf{p}^0)  \rangle\, .
\end{equation}
By imposing 
$\langle \Psi_T (\mathbf{p}^0)|\bar{\Psi}_T (\mathbf{p}^0) \rangle=1$,
we ensure that 
\begin{align}
 |\bar{\Psi}^i_T  (\mathbf{p}^0)  \rangle &
 =\frac{\partial | \bar{\Psi}_T(\mathbf{p}) \rangle}{\partial p_i}
  \Big|_{\mathbf{p}=\mathbf{p}^0}\nonumber\\
 &= |\Psi_T^i(\mathbf{p}^0) \rangle - S_{0i}|\Psi_T(\mathbf{p}^0) \rangle
\end{align}
are orthogonal to $|\Psi_T (\mathbf{p}^0) \rangle$. 
In the last equation we have introduced
\begin{equation}
|\Psi_T^i (\mathbf{p}^0 ) \rangle = \frac{ \partial |\Psi_T(\mathbf{p})\rangle}
{\partial p_i}\Big|_{\mathbf{p}=\mathbf{p}^0}
\end{equation}
for the first derivative with respect to the $i$-th 
parameter, and the overlap matrix is defined by 
$S_{0i}=\langle \Psi_T (\mathbf{p}^0) | \Psi_T^i  (\mathbf{p}^0)\rangle$.
The 
expectation value of the energy on the linear wave function is defined as
\begin{equation}
E_\text{lin}(\mathbf{p})\equiv \frac{\langle\bar{\Psi}_T^\text{lin}(\mathbf{p})|
H  |\bar{\Psi}_T^\text{lin}(\mathbf{p})\rangle}
{\langle\bar{\Psi}_T^\text{lin}(\mathbf{p})|\bar{\Psi}_T^\text{lin}(\mathbf{p})
\rangle} 
\, .
\end{equation}
The variation $\Delta\bar{\mathbf{p}}$ of the parameters that minimizes 
the energy, $\nabla_\mathbf{p} E_\text{lin}(\mathbf{p})=0$, 
corresponds to the lowest eigenvalue solution of the generalized eigenvalue 
equation
\begin{equation}
\bar{H} \, \Delta \mathbf{p} = \Delta E \, \bar{S} \, \Delta \mathbf{p}\, ,
\label{eq:lin_sys}
\end{equation}
where $\bar{H}$ and $\bar{S}$ are the Hamiltonian and overlap matrices in 
the $(N_p+1)$-dimensional basis defined by 
$\{|\bar{\Psi}_T(\mathbf{p}^0)\rangle, |\bar{\Psi}^1_T(\mathbf{p}^0)\rangle, 
\dots, |\bar{\Psi}^{N_p}_T(\mathbf{p}^0)\rangle\}$. The 
authors of Ref.~\cite{Umrigar:2005} have shown that writing the expectation 
values of these matrix elements in terms of covariances allows 
us to keep their statistical error under control even when they are estimated 
over a relatively small Monte Carlo sample. However, since in AFDMC the 
derivatives of the wave function with respect to the orbital variational 
parameters are in general complex, we generalized the expressions for the 
estimators reported in the appendix of Ref.~\cite{Umrigar:2005}. 

For a finite sample size the matrix $\bar{H}$ can be
ill-conditioned, spoiling therefore the numerical inversion needed to solve 
the eigenvalue problem. A practical procedure to stabilize the algorithm is 
to add a small positive constant 
$\epsilon$ to the diagonal matrix elements of $\bar{H}$ 
except for the first one, 
$\bar{H}_{ij} \to \bar{H}_{ij} + \epsilon (1-\delta_{i0})\delta_{ij}$. 
This procedure reduces the length of $\Delta\bar{\mathbf{p}}$ and rotates it 
towards the 
steepest-descent direction.

It has to be noted that if the wave function depends linearly upon the 
variational parameters, the algorithm converges in just one iteration. 
However, in our case strong nonlinearities in the variational parameters 
make, in some instances, $|\bar{\Psi}_T^\text{lin}(\mathbf{p})\rangle$ 
significantly different from 
$|\bar{\Psi}_T(\mathbf{p}^0+\Delta\mathbf{p})\rangle$.
Accounting for the quadratic term in the expansion as in the 
Newton method~\cite{Lee:2005,Umrigar:2005} would alleviate the problem, 
at the expense of having to estimate also the 
Hessian of the wave function with respect to the variational parameters. 
An alternative strategy consists in taking advantage of the arbitrariness 
of the wave-function normalization to improve on the 
convergence by a suitable rescaling of the parameter 
variation~\cite{Umrigar:2005,Toulouse:2007}. We found that this procedure was 
not sufficient to guarantee the stability of the minimization procedure. 
For this reason we have implemented the following heuristic procedure.  
For a given value of $\epsilon$, Eq.~\eqref{eq:lin_sys} is solved. 
If the linear variation of 
the wave function for $\mathbf{p}=\mathbf{p}^0 +\Delta {\mathbf{p}}$ is small,  
\begin{equation}
\frac{| \bar{\Psi}_T^\text{lin}(\mathbf{p})|^2}{| \bar{\Psi}_T(\mathbf{p}^0)|^2}=
1 + \sum_{i,j=1}^{N_p} \bar{S}_{ij}\Delta{p}^{\, i} \Delta{p}^{\, j}\le \delta
\, ,
\label{eq:lin_var}
\end{equation}
a short correlated run is performed in which the energy expectation value
\begin{equation}
E(\mathbf{p})\equiv \frac{\langle \bar{\Psi}_T(\mathbf{p} ) | H  
|\bar{\Psi}_T(\mathbf{p} ) \rangle }{\langle  \bar{\Psi}_T (\mathbf{p} ) 
|\bar{\Psi}_T(\mathbf{p} ) \rangle}
\end{equation}
is estimated along with the full variation of the wave function for a set of 
possible values of $\epsilon$ (in our case $\approx100$ values are considered).
The optimal $\epsilon$ is chosen 
so as to minimize $E(\bar{\mathbf{p}})$ provided that 
\begin{equation}
\frac{|\bar{\Psi}_T(\bar{\mathbf{p}})|^2}{|\bar{\Psi}_T(\mathbf{p}^0)|^2}  
\le \delta
\, .
\label{eq:full_var}
\end{equation}
Note that, at variance with the previous expression, here in the numerator we 
have the full 
wave function instead of its linearized approximation. In the (rare) cases where
no acceptable value of $\epsilon$ is found due to possibly large statistical
fluctuations in the VMC estimators, we perform an additional run adopting the 
previous parameter set and a new optimization is attempted. 
In our experience, this procedure proved extremely robust.

The chief advantage of the additional 
constraint is that it suppresses the 
potential instabilities caused by the nonlinear dependence of the wave function
on the variational parameters. When using the ``standard'' version of the LM, 
there were instances in which, despite the variation of the linear wave 
function 
being well below the threshold of Eq. (\ref{eq:lin_var}), 
the full wave function fluctuated significantly more, preventing the 
convergence of the minimization algorithm.  As for the 
wave-function variation, we found that choosing $\delta=0.2$ guarantees a fast 
and stable convergence. 

The two-body Jastrow correlation $f(r_{ij})$ is written in terms of cubic 
splines, characterized by a smooth first derivative and continuous second 
derivative. The adjustable parameters to be optimized are the ``knots'' 
of the spline, which are simply the values of the Jastrow function at the 
grid points, and the value of the first derivative at $r_{ij}=0$. 
Analogous 
parametrizations are adopted for $u(r_{ij})$ and $R_{n\ell}(r_i)$. 
In the $^4$He case we used six variational parameters for $f(r)$, $u(r)$,
and for the radial orbital functions $R_{n\ell}(r)$. 
This allowed enough flexibility for the variational energies to be very close 
to the one obtained performing the imaginary-time diffusion 
for all 
values of the cutoff and of the pion mass. 
On the other hand, to allow 
for 
an emerging cluster structure, for the $^{16}$O wave function we used 30 
parameters for the two- and three-body Jastrow correlations and 15 parameters 
for each of the $R_{n\ell}(r)$. 

\begin{center}
\begin{figure}[tb]
\includegraphics[width=9cm]{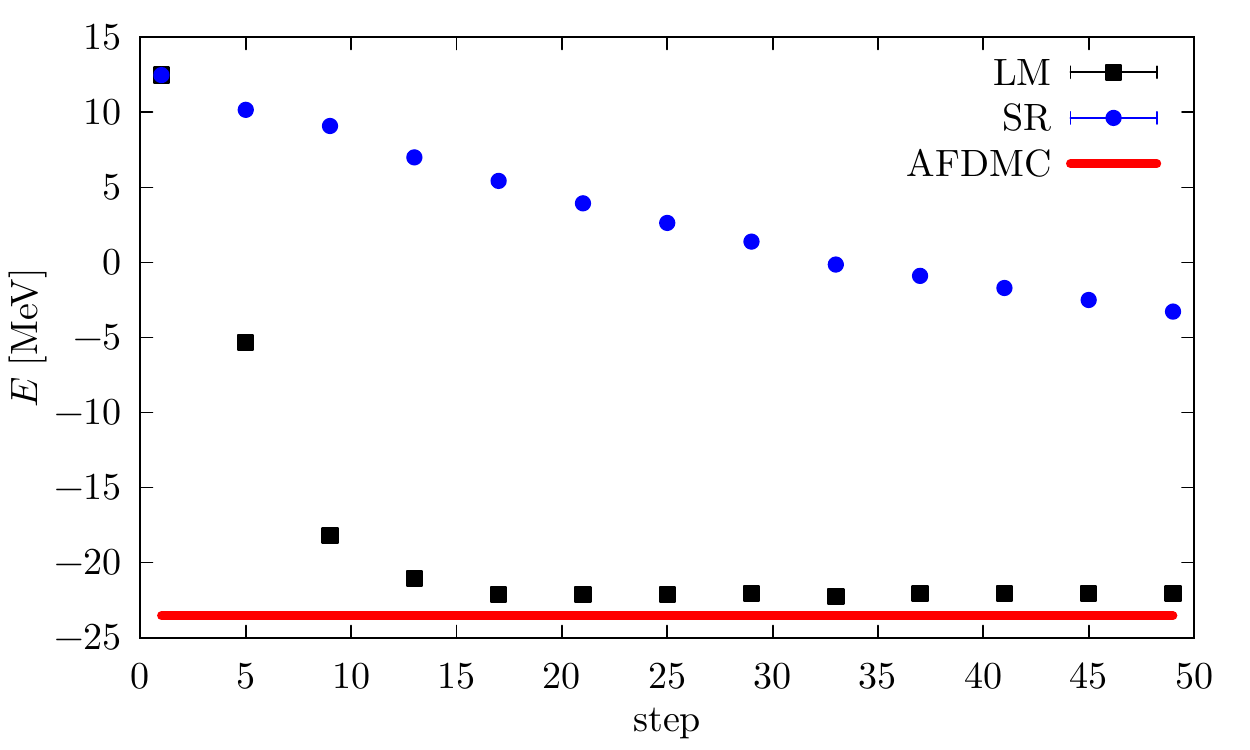}
\caption{(Color online) Convergence pattern of the 
$^4$He variational energy at physical pion mass and $\Lambda=4$ fm~$^{-1}$
as a function of the number of optimization steps for
the SR method (black squares) and the LM (blues circles). For comparison, 
the red line indicates the AFDMC result.}
\label{fig:he4_lm_sr}
\end{figure}
\end{center}

The LM exhibits a much faster convergence pattern than the SR, previously used 
in AFDMC. In Fig. \ref{fig:he4_lm_sr}, we show the $^4$He variational energy 
obtained for physical pion mass and $\Lambda=4$ fm~$^{-1}$ as a function of the 
number of 
optimization steps for both SR and LM.
While the LM takes only $\simeq15$ steps to converge, the SR is much slower; 
after 50 steps the energy is still much above the asymptotic limit. 
We have observed analogous 
behavior for other values of the cutoff and the pion mass. In the $^{16}$O case,
the 
improvement of the LM with respect to the SR is even more dramatic due 
to the clustering of the wave function, which will be discussed in detail 
in the following. 

\section{Results}
\label{sec:results}

With LO \eftnopi~LECs determined from experiment or LQCD calculations,
predictions can be made with AFDMC 
for the binding energies of $^4$He and $^{16}$O.

In Table \ref{tab:he4_binding} we report 
results of $^4$He 
energies for all the values of the cutoff and of the pion mass
we considered. Despite the different parametrization of the variational 
wave functions, the results are in very good agreement with those reported in 
Ref. \cite{Barnea:2013uqa}, where a simplified version of the variational 
wave function was used because the LM had not been introduced yet. 
For most cutoff values, our results also agree with those of 
Ref. \cite{Kirscher:2015yda}, which were obtained with the
Resonating-Group and Hyperspherical-Harmonic methods.
For $\Lambda\leq 6~\text{fm}^{-1}$, the QMC results differ by less than 
0.1~MeV from Ref.~\cite{Kirscher:2015yda}, while for $\Lambda=8~\text{fm}^{-1}$
the QMC method binds $^4$He more deeply by more than 1~MeV.
In consequence, the extrapolated asymptotic values differ.
Our results display a better convergence pattern with the cutoff.
At the physical pion mass and with the same input observables, 
our highest-cutoff result is in good
agreement with the highest-cutoff result
(cutoff values in the range $8-10$ fm$^{-1}$, but a different regulator
function) of Ref. \cite{Platter:2004zs}.

\begin{table}[tb]
\renewcommand{\arraystretch}{1.2}
\small
\begin{center}
\begin{tabular}{ c | c c c }
\hline
\hline
$\Lambda$ & $m_\pi=140$ MeV  &  $m_\pi=510$ MeV & $m_\pi=805$ MeV \\
\hline
$2$ fm$^{-1}$ & $-23.17 \pm 0.02$  &   $-31.15 \pm 0.02$  & $-88.09 \pm 0.01$ \\
$4$ fm$^{-1}$ & $-23.63 \pm 0.03$  &   $-34.88 \pm 0.03$  & $-91.40 \pm 0.03$ \\
$6$ fm$^{-1}$ & $-25.06 \pm 0.02$  &   $-36.89 \pm 0.02$  & $-96.97 \pm 0.01$ \\
$8$ fm$^{-1}$ & $-26.04 \pm 0.05$  &   $-37.65 \pm 0.03$  & $-101.72 \pm 0.03$
\phantom{a} \\
$\rightarrow\infty$ & $-30_{\pm 2\,\text{(stat)}}^{\pm 0.3\,\text{(sys)}}$ 
& $-39 _{\pm 2\,\text{(stat)}}^{\pm 1\,\text{(sys)}}$ 
& $-124_{\pm 1\,\text{(stat)}}^{\pm 3\,\text{(sys)}}$  \\
\hline
Exp. & $-28.30$ & -- & -- \\
LQCD & -- & $-43.0 \pm 14.4$ & $-107.0 \pm 24.2$ \\
\hline
\end{tabular}
\caption{
$^4$He energy for different values of the pion mass $m_\pi$ and the cutoff 
$\Lambda$, compared to experiment and LQCD calculations
\cite{Beane:2012vq,Yamazaki:2012hi}.
See main text and appendix for details on 
errors and extrapolations.}
\label{tab:he4_binding}
\end{center}
\end{table}

We found that an expansion 
of the type \eqref{eq:cut_off_extrapolation} up to  $1/\Lambda^2$ suffices to 
extrapolate the $^4$He energies for $m_\pi=140$ MeV,
since the addition of a cubic term changes neither the extrapolated value 
nor the best-fit coefficients.
For the unphysical pion masses, the usage of the smallest cutoff is
questionable because $\Lambda=2$ fm$^{-1}$ cuts off momentum modes
below the pion mass. 
We thus extrapolate the values appearing in the tables 
with the quadratic expansion in Eq.~\eqref{eq:cut_off_extrapolation} 
but without the result at $\Lambda=2$ fm$^{-1}$. 
In all cases, we perform fits with and without the $\Lambda=2$ fm$^{-1}$
results to estimate the systematic extrapolation error.
The procedure adopted for the systematic and statistical
errors quoted throughout this paper is detailed in the appendix.

It has to be remarked that this cutoff sensitivity study does not account for 
the EFT truncation 
error.
Using cutoff variation from cutoff values somewhat larger than
the pion mass, for example from 
$\Lambda = 2$ fm$^{-1}$, 4 fm$^{-1}$, and 6 fm$^{-1}$ for 
$m_\pi=140$, $510$, and $805$ MeV, 
we might estimate the error as $\pm 7$, $\pm 4$, and $\pm 30$,
respectively. Except for the intermediate pion mass,
this is consistent with the rougher dimensional-analysis estimate 
$Q_A/M\sim 0.3$. 
In any case, we expect the truncation error to dominate 
over the statistical and extrapolation errors.

Given the convergence of the $^4$He binding energy with increasing cutoff, 
we confirm that, for both physical \cite{Platter:2004zs} and 
unphysical \cite{Barnea:2013uqa,Kirscher:2015yda} pion masses,
LO \eftnopi~is renormalized correctly without
the need for a four-nucleon interaction.
In the physical case,
the binding energy is underestimated for all 
values of the cutoff we considered,
but the extrapolated value is in agreement with experiment even if we
neglect the truncation error. Of course when the latter is taken into account
we must conclude that such a good agreement is somewhat fortuitous.
We expect NLO corrections, including Coulomb and 
two-nucleon effective-range corrections, to change the result by a few MeV.
For $m_\pi=510$ MeV and $m_\pi=805$ MeV, our results reproduce
LQCD predictions (where Coulomb is absent)
within the measurement error over the entire cutoff range. 
As pointed out in Refs. \cite{Barnea:2013uqa,Kirscher:2015yda},
this is a non-trivial consistency check:
if either LQCD data or \eftnopi~were too wrong, one would expect no such 
agreement.
However, LQCD uncertainties are too large at this point
for us to draw a very strong conclusion. 

It is interesting to study the cutoff dependence of the root-mean-square (rms) 
point-nucleon radius $\sqrt{\smash[b]{\langle r^{2}_\text{pt} \rangle}}$ and 
the single-nucleon point density $\rho_\text{pt}(r)$. These quantities 
are related to the charge density, which can be extracted from electron-nucleus
scattering data, but are not observable themselves:
few-body currents and single-nucleon electromagnetic form factors have to be 
accounted for. Still, one can gain some insight into the features
of the ground-state wave function by comparing results
at different pion masses and cutoffs.
Since neither $\sqrt{\smash[b]{\langle r^{2}_\text{pt} \rangle}}$ nor 
$\rho_\text{pt}(r)$ commute with the Hamiltonian, the desired expectation 
values on the ground-state wave function are computed by means of 
``mixed'' matrix elements
\begin{equation}
\langle \Psi_ 0 | \mathcal{O} | \Psi_0 \rangle \approx 
2 \langle \Psi_T | \mathcal{O} | \Psi_0 \rangle 
- \langle \Psi_T | \mathcal{O} | \Psi_T \rangle\, .
\end{equation}
In the above equation $|\Psi_0 \rangle$ is the imaginary-time evolved state of 
Eq. (\ref{eq:tau_ev}), while  $|\Psi_T\rangle$ is the trial wave function
constructed as in Eq. (\ref{eq:trial}). 

The results for the 
point-proton radius of $^4$He are reported in Table \ref{tab:he4_radii}. 
(Since Coulomb is absent in our calculation, the point-nucleon 
and point-proton radii are the same.)
In the physical 
case, the calculated radius is much smaller than the 
empirical value --- that is, the value extracted from
the experimental data of Ref. \cite{Ottermann:1985km} accounting for
the nucleon size, but neglecting meson-exchange currents. 
A similar result, $\sqrt{\smash[b]{\langle r^{2}_\text{pt} \rangle}}\approx 1$ fm
was obtained by the 
authors of Ref. \cite{Lynn:2014zia} using a local form of a
chiral interaction.
NLO and N$^2$LO potentials in 
a chiral expansion based on naive dimensional analysis 
\cite{Bedaque:2002mn,Epelbaum:2008ga,Machleidt:2011zz}
bring theory into much closer agreement with the 
empirical value. 
Hence, sub-leading terms in the \eftnopi~expansion could play a relevant role,
at least for physical values of the pion mass.

\begin{center}
\begin{figure}[t]
\includegraphics[width=9cm]{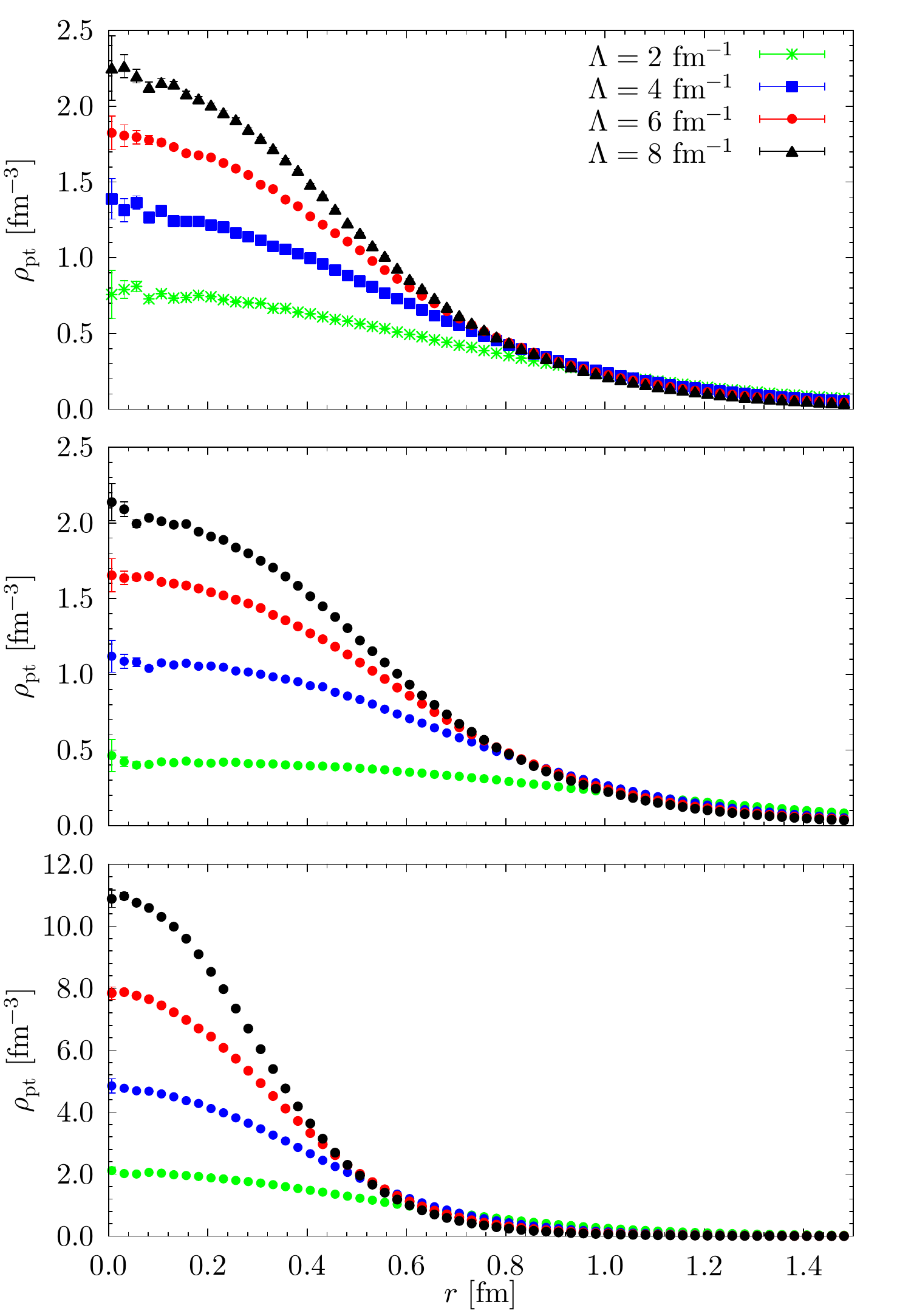}
\caption{(Color online) 
$^4$He single-nucleon point density 
for $m_\pi=140$ MeV (upper panel), $m_\pi=510$ MeV 
(middle panel), and $m_\pi=805$ MeV (lower panel),
at different values of the cutoff $\Lambda$.}
\label{fig:he4_rho}
\end{figure}
\end{center}

For unphysically large pion masses, where 
\eftnopi~is supposed to exhibit a faster convergence, the 
point-proton radius is smaller than at $m_\pi=140$ MeV. 
The value obtained for $m_\pi=510$ MeV indicates a spatial extent similar to the
physical one, while $^4$He at $m_\pi=805$ MeV, in comparison, seems to be a 
much more compact object.
This is consistent with the behavior of the single-nucleon point density,
$\rho_{\text{pt}}$, displayed in Fig. \ref{fig:he4_rho}. 
For all cutoff values, the 
density corresponding to $m_\pi=805$ MeV 
is appreciably narrower than 
that computed for $m_\pi=510$ MeV 
or $m_\pi=140$ MeV. Focusing on 
$\Lambda=8$ fm$^{-1}$,
$\rho_{\text{pt}}$ has a maximum value of $11.0$ fm$^{-3}$ for $m_\pi=805$ MeV, 
while in the $m_\pi=510$ MeV and $m_\pi=140$ MeV cases the maximum values are 
$2.1$ fm$^{-3}$ and $2.2$ fm$^{-3}$, respectively.

\begin{table}[tb]
\begin{center}
\renewcommand{\arraystretch}{1.2}
\small
\begin{tabular}{c | c c c }
\hline
\hline
$\Lambda$ & $m_\pi=140$ MeV  &  $m_\pi=510$ MeV & $m_\pi=805$ MeV \\
\hline
$ 2$ fm$^{-1}$ & $1.374 \pm.0.004 $  & $1.482 \pm 0.003$  & $0.898 \pm 0.001$ \\
$ 4$ fm$^{-1}$ & $1.203 \pm 0.004$   & $1.133 \pm 0.003$  & $0.699 \pm 0.001$ \\
$ 6$ fm$^{-1}$ & $1.109 \pm 0.003$   & $1.035 \pm 0.002$  & $0.609 \pm 0.001$ \\
$ 8$ fm$^{-1}$ & $1.054 \pm 0.003$   & $0.976 \pm 0.001$  & $0.542 \pm 0.001$ \\
$\rightarrow\infty$ & $ 0.9_{\pm 0.2\,\text{(stat)}}^{\pm 0.008\,\text{(sys)}}$ 
& $0.8_{\pm 0.1\,\text{(stat)}}^{\pm 0.04\,\text{(sys)}}$ 
& $0.25_{\pm 0.06\,\text{(stat)}}^{\pm 0.05\,\text{(sys)}}$ \\
\hline
``Exp.'' & $1.45 $ & -- & -- \\
\hline
\end{tabular}
\caption{
$^4$He point-proton radius for different values of the pion mass $m_\pi$ 
and the cutoff $\Lambda$, compared to the empirical value extracted
from Ref. \cite{Ottermann:1985km} accounting for the finite nucleon size. 
See main text and appendix for details on errors and extrapolations.}
\label{tab:he4_radii}
\end{center}
\end{table}

The similarity between $^4$He ground-state 
properties at $m_\pi=510$ MeV and those at the physical pion mass
exists despite differences in the structure of lighter systems.
If confirmed for other properties of $^4$He and 
heavier nuclei, this semblance would
mean that simulations at intermediate pion masses could provide useful insights 
into the physical world while saving substantial computational resources. 

In Table \ref{tab:o16_binding} the 
$^{16}$O ground-state energies are reported for the same 
pion-mass and cutoff values considered for $^4$He. 
A striking feature is that $^{16}$O is not stable against 
breakup into four $^{4}$He clusters in almost all the cases, 
the only exception occurring for $m_\pi=140$ MeV and $\Lambda=2$ fm$^{-1}$,
where $^{16}$O is $4.5$ MeV more bound than four $^{4}$He nuclei.
In the other cases we miss the four-$^{4}$He threshold by about $5$ MeV, which 
is beyond our statistical errors and reveals a lower bound
on the systematic error of our QMC method. 

\begin{table}
\renewcommand{\arraystretch}{1.2}
\small
\begin{center}
\begin{tabular}{c | c c c }
\hline
\hline
$\Lambda$ & $m_\pi=140$ MeV  &  $m_\pi=510$ MeV & $m_\pi=805$ MeV  \\
\hline
$2$ fm$^{-1}$ & $-97.19 \pm 0.06 $ & $-116.59 \pm 0.08$ & $-350.69 \pm 0.05$ \\
$4$ fm$^{-1}$ & $-92.23 \pm 0.14 $ & $-137.15 \pm 0.15$ & $-362.92 \pm 0.07$ \\
$6$ fm$^{-1}$ & $-97.51 \pm 0.14 $ & $-143.84 \pm 0.17$ & $-382.17 \pm 0.25$ \\
$8$ fm$^{-1}$ & $-100.97 \pm 0.20$ & $-146.37 \pm 0.27$ & $-402.24 \pm 0.39$ \\
$\rightarrow\infty$ & $ -115_{\pm 8\,\text{(stat)}}^{\pm 1\,\text{(sys)}}$ 
& $-151_{\pm 10\,\text{(stat)}}^{\pm 2\,\text{(sys)}}$ 
& $-504_{\pm 12\,\text{(stat)}}^{\pm 20\,\text{(sys)}}$ \\
\hline
Exp. & $-127.62$ & -- & -- \\
\hline
\end{tabular}
\caption{
$^{16}$O energy for different values of the pion mass $m_\pi$ and the cutoff 
$\Lambda$, compared with experiment. (No LQCD data exist for this nucleus.)
See main text and appendix for details on errors and extrapolations.}
\label{tab:o16_binding}
\end{center}
\end{table}

Even considering only statistical and extrapolation errors
the asymptotic values of the $^{16}$O energy cannot be separated from
the four-$^{4}$He threshold.
The proximity of the threshold suggests that the structure of
our $^{16}$O should be clustered.
Indeed,
despite no explicit clustering 
being enforced in the trial wave function, the highly efficient optimization 
procedure arranges the two- and three-body Jastrow correlations, as well as 
the orbital radial functions, in such a way 
as to favor configurations characterized by four independent $^4$He clusters. 

\begin{center}
\begin{figure}[h!]
\includegraphics[width=9cm]{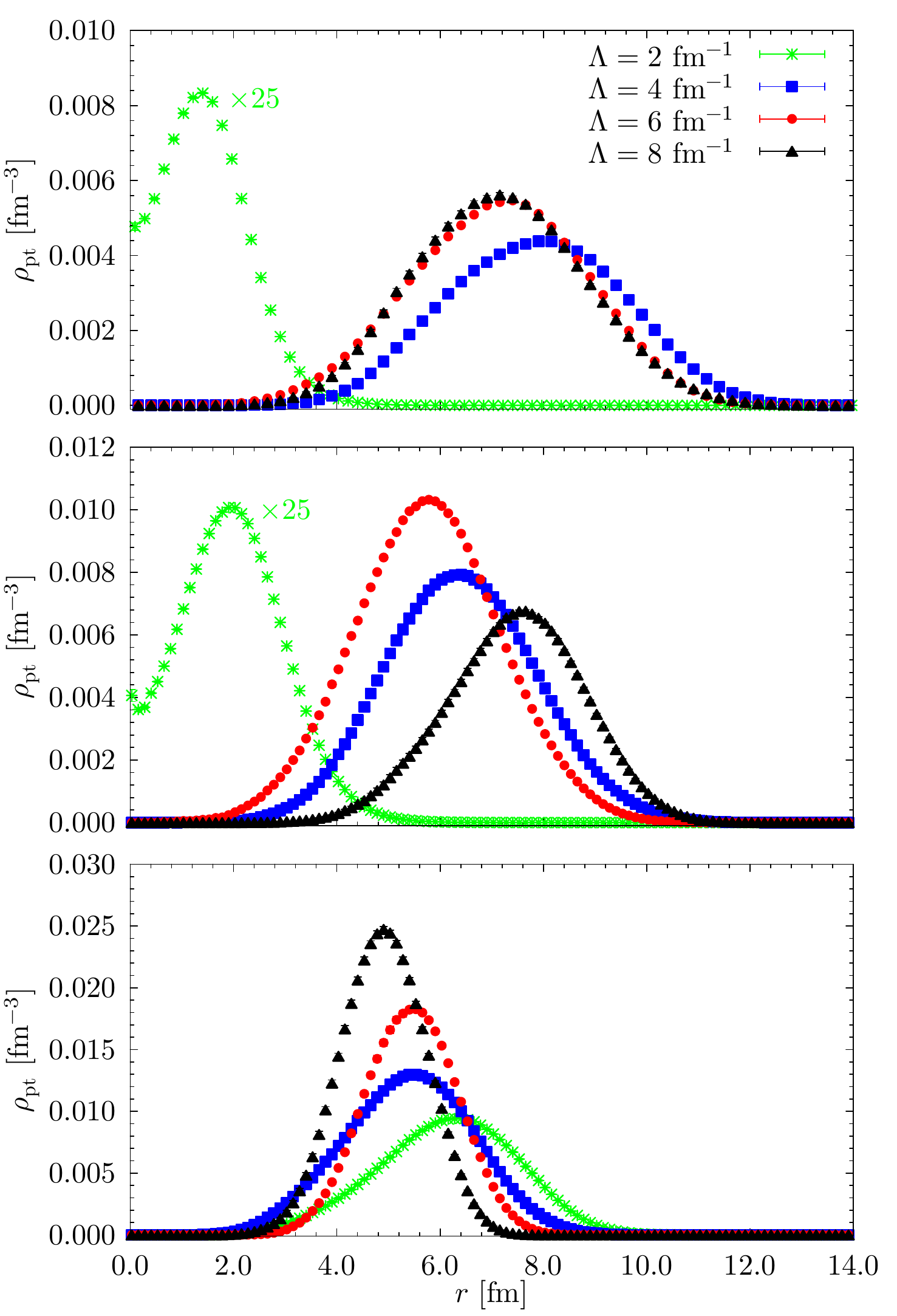}
\caption{(Color online) 
$^{16}$O single-nucleon point density 
for $m_\pi=140$ MeV (upper panel), $m_\pi=510$ MeV 
(middle panel), and $m_\pi=805$ MeV (lower panel), 
at different values of the cutoff $\Lambda$.}
\label{fig:o16_rho}
\end{figure}
\end{center}

The single-proton density profiles displayed in Fig. \ref{fig:o16_rho} indicate
that only for $\Lambda=2$ fm$^{-1}$ with $m_\pi=140$ MeV and $m_\pi=510$ MeV 
are the nucleons
distributed according to the classic picture of a bound wave function. 
For all the other combinations of pion masses and cutoffs, nucleons are pushed 
away from the center of the nucleus, which is basically empty 
--- the density at the origin is a minuscule fraction of the peak ---
until $\simeq 2$ fm from the center of mass. The erratic behavior 
of the peak position of the density profiles as a function of the cutoff 
has to be ascribed to the fact that the relative position of the four $^{4}$He 
clusters is practically unaffected by the cutoff value. In fact, once the 
clusters are sufficiently apart, a landscape of degenerate minima in the 
variational energy emerges. Hence, the single-proton densities correspond 
to wave functions that, despite potentially significantly different, lead 
to almost identical variational energies. 
In contrast, the width of the peaks decreases with increasing cutoff
in step with the shrinking of 
the individual
$^{4}$He clusters
reported in Table \ref{tab:he4_radii}.

The analysis of the proton densities alone does not suffice to support the 
claim of clustering. Another indication of clusterization comes from comparing 
the expectation values of the nuclear potentials evaluated in the 
ground states of $^{16}$O and $^{4}$He. For instance, in the $m_\pi=140$ MeV 
and $\Lambda=8$ fm$^{-1}$ case it turns out that the 
expectation values of the $^{16}$O two- and three-body potentials 
are $\simeq 4.05$ and $\simeq 4.16$ times larger than the corresponding 
values for $^4$He. The same pattern is observed for all the combinations of 
pion mass and cutoff, except for $\Lambda=2$ fm$^{-1}$ with $m_\pi=140$ MeV and 
$m_\pi=510$ MeV. In particular, for $\Lambda=2$ fm$^{-1}$ and $m_\pi=140$ MeV, 
the 
expectation values of the two- and three-body potentials in $^{16}$O are 
$\simeq 4.65$ and $\simeq 6.14$ times larger than 
in $^4$He. This 
difference is a consequence of the fact that the number of interacting 
pairs and triplets is larger when clusterization does not take place. 

\begin{center}
\begin{figure}[t]
\includegraphics[width=10cm]{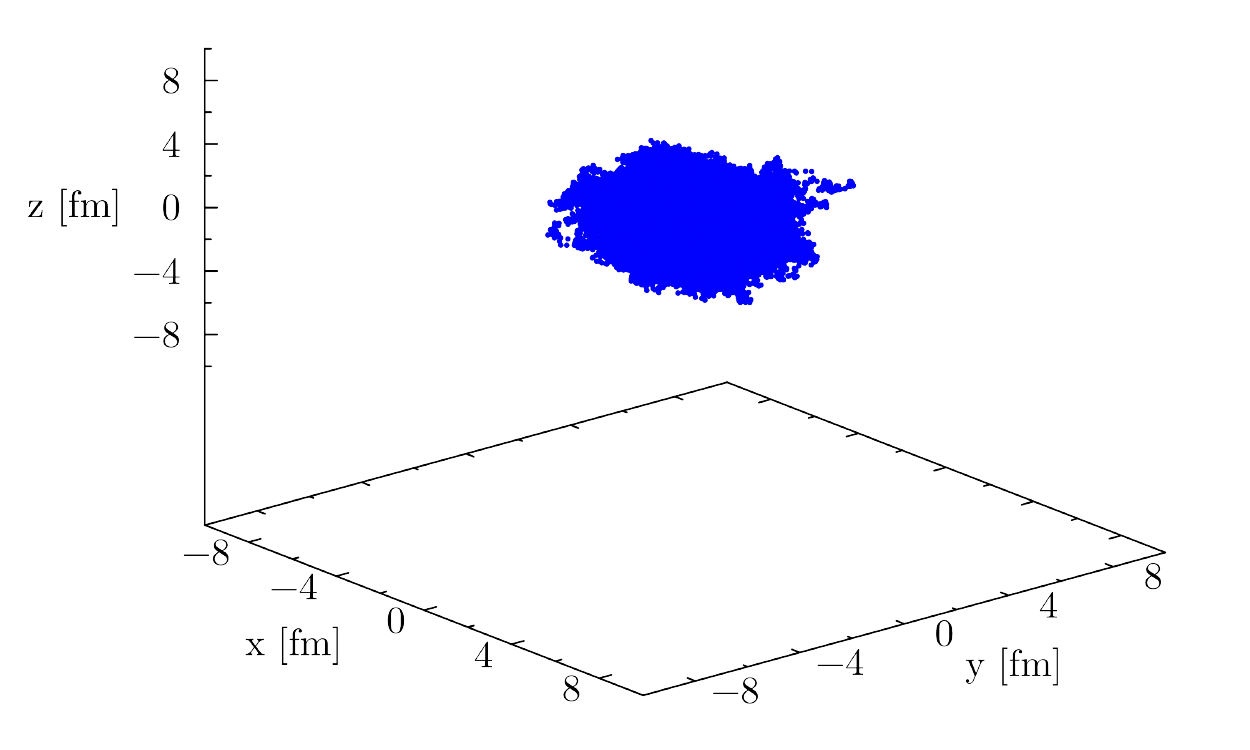}
\includegraphics[width=10cm]{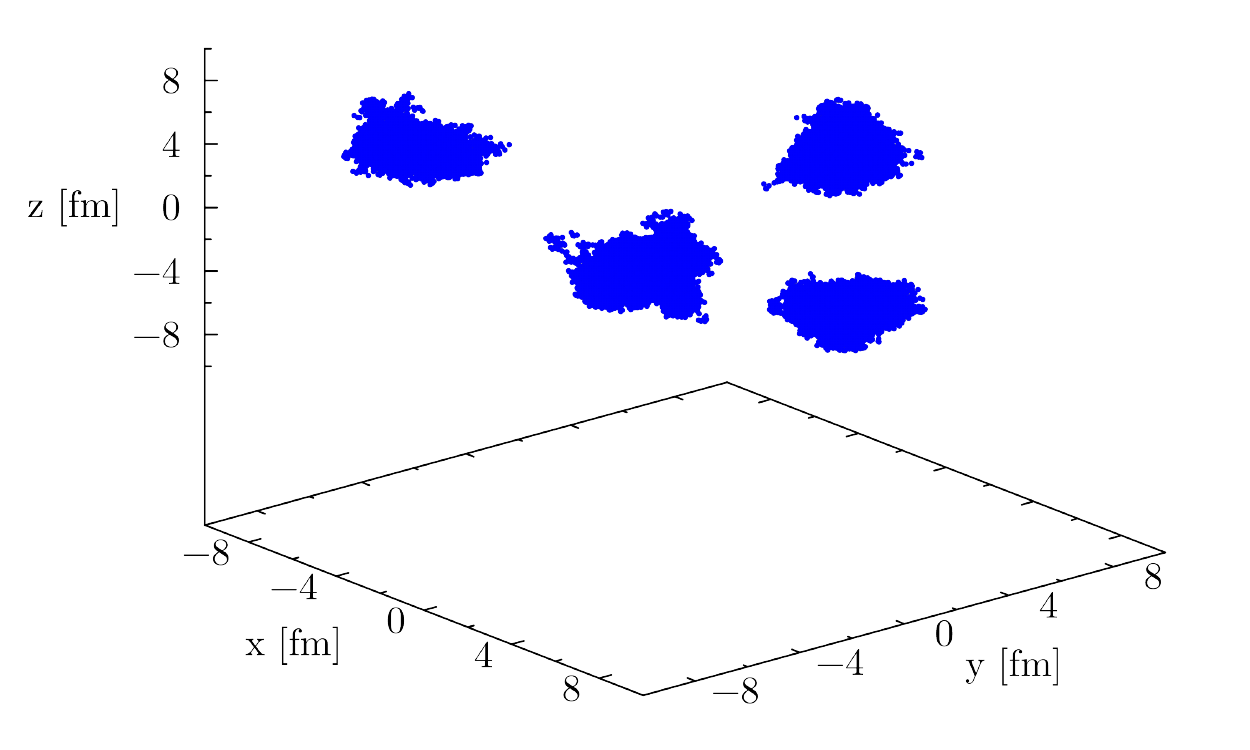}
\caption{(Color online) Imaginary-time diffusion 
with time step $\Delta\tau=0.125$ MeV$^{-1}$ of a single walker 
for $m_\pi=140$ MeV,
at $\Lambda=2$ fm$^{-1}$ (upper panel) and $\Lambda=8$ fm$^{-1}$ 
(lower panel).}
\label{fig:cluster}
\end{figure}
\end{center}

To better visualize the clusterization of the wave function, in 
Fig. \ref{fig:cluster} we display the position of the nucleons following the 
propagation of a single walker for 5000 imaginary-time
steps, corresponding to $\Delta\tau=0.125$ MeV$^{-1}$, printed every 10 steps. 
In the upper panel, 
concerning $m_\pi=140$ MeV and $\Lambda=2$ 
fm$^{-1}$, 
nucleons are not organized in clusters. In fact, during the imaginary-time 
propagation they diffuse in the region in which the corresponding 
single-nucleon density of Fig. \ref{fig:o16_rho} does not vanish. 
A completely different scenario takes place 
at the same pion mass when 
$\Lambda=8$ fm$^{-1}$:  
the nucleons forming the four $^{4}$He clusters remain close to the 
corresponding centers of mass during the entire imaginary-time propagation. 
This is 
clear evidence of clustering. It has to be noted that the relative position 
of the four 
clusters is not a tetrahedron. To prove this, for each configuration we 
computed the 
moment-of-inertia matrix as in Ref.~\cite{Wiringa:2000gb}. 
If the $^{4}$He clusters were positioned at the 
vertices of a tetrahedron,  
diagonalization would yield only 
two independent eigenvalues. 
Instead, we found three distinct eigenvalues, corresponding to an ellipsoid 
--- yet another indication of the absence of interactions among nucleons 
belonging to different  $^{4}$He clusters.

The non-clustered states at
$\Lambda=2~$fm$^{-1}$ for $m_\pi=510$ MeV and $m_\pi=140$ MeV
deserve further comment.
The state at $m_\pi=510$ MeV
stands in contrast to the other states found above threshold
whose structure is clustered.
We interpret this 
as an artifact of the numerical method, since 
a perfect optimization procedure should have produced a clustered 
structure resembling the lower-energy state with four free $^{4}$He.
While there is no signal of $^{16}$O stability above the physical
pion mass,
the state at $m_\pi=140$ MeV is certainly stable at the lowest
cutoff, that is, when the interaction has the longest range.
On this basis, one might speculate that at some pion mass
above the physical one a transition from
a non-clustered to a clustered state is expected.
However, such a conclusion cannot be drawn 
until higher-order calculations in \eftnopi~--- which will capture 
finer effects from pion exchange such as the tensor force at N$^2$LO --- 
are available.

The smaller relative size of the model space leads to more
modest signs of cutoff convergence for $^{16}$O than $^4$He, which are
reflected in larger extrapolation errors, especially at $m_\pi=805$ MeV.
At physical pion mass, the central value of the extrapolated total energy 
is only 10\% off experiment, 
which can be bridged by statistical
and extrapolation errors. 
This difference is small compared to 
the expected truncation error, $\sim 30\%$. 
If there is a low-lying resonant or virtual state of $^{4}$He nuclei
at LO in \eftnopi~--- note that our analysis does neither preclude nor 
identify such a state --- it is possible that the (perturbative) inclusion 
of higher-order terms up to N$^2$LO will move the $^{16}$O energy 
sufficiently for stability with respect to four $^{4}$He clusters.
 
For unphysical pion mass, our results can be seen as an extension
of LQCD to medium-mass nuclei, with no further assumptions 
about the QCD dynamics.
In this case, a determination of the relative position of the 
four-$^{4}$He threshold 
would further require much increased accuracy in the $A=2,3$ LQCD data 
that we use as input. 

\section{Conclusions}
\label{sec:conclusions}

This paper represents the first application of the effective-field-theory
formalism, as developed for small nuclei without explicit pions, to a
relatively heavy object, $^{16}$O. 
We employed contact potentials which represent 
the leading order of a systematic expansion of QCD.
This enabled us to analyze physical nucleons as well as 
simulated scenarios with increased quark masses.

To overcome the peculiar challenges associated with the
solution of the Schr\"odinger equation,
we have improved AFDMC by introducing a new 
optimization protocol of the many-body wave function to be employed in the 
variational stage of the calculation. The scheme we 
propose is an extension of the linear method and provides a much faster 
convergence in 
parameter space compared to 
stochastic reconfiguration, previously adopted in nuclear QMC calculations. 
Such accurate trial wave function is the starting point of the 
imaginary-time projection in AFDMC, 
which filters out the ``exact'' ground state of the Hamiltonian. 
This algorithm was used to predict not only 
ground-state energies,
but also radii, densities, and particle distributions.

Our results for the $^4$He binding energy are in agreement with previous
findings, including the renormalizability of the four-nucleon system
in \eftnopi~without a LO four-body force. 
In particular, at physical pion mass the energy agrees with experiment
within theoretical uncertainties. Moreover, the calculated
point-nucleon radii and single-particle densities reveal 
a $^4$He structure at $m_\pi=510$ MeV similar
to that at physical pion mass.

With this successful benchmark, we extended the calculations to $^{16}$O, 
obtaining extrapolated values for the $^{16}$O energy at all pion masses
which are indistinguishable from the respective four-$^{4}$He threshold, even
considering only the smaller statistical and extrapolation errors.
In fact, for almost all cutoffs and pion masses we considered,
$^{16}$O is unstable with respect to 
break-up into four $^{4}$He nuclei. 
Our calculation of the $^{16}$O energy is the 
first time LQCD data are extended to the medium-mass region
in a model-independent way \footnote{As this manuscript was being concluded,
a calculation of doubly magic nuclei appeared \cite{McIlroy:2017ssf},
where a two-body potential model obtained from
LQCD data at $m_\pi=469$ MeV was solved with the Self-Consistent Green's 
Function method. The widely different input data and method translate
into much smaller $^4$He and $^{16}$O energies than our results. 
No clear sign is found 
for a $^{16}$O state below the $^4$He threshold.}.

Interestingly, $m_\pi=140$ MeV and $\Lambda=2$ fm$^{-1}$ is the 
only parameter set which yields a 
stable $^{16}$O. This suggests that 
the long-range structure of the interaction is deficient at larger cutoff values
and might have to be corrected, {\it e.g.} via one-pion exchange,
to guarantee the binding of heavier nuclei {\it at LO}.
Alternatively,
within a pionless framework, higher-order terms could act as perturbations 
to move $^{16}$O with respect to the four-$^4$He threshold.
At physical pion mass the central value of the
total energy is just about 10\% off experiment.
This is only slightly larger than the 
statistical and extrapolation errors, and well within
the $\sim 30\%$ truncation error estimate. 
We cannot exclude the possibility
that agreement with data will improve with order.
A comprehensive study of the various subsystems of $^{16}$O
--- for example, $^{12}$C, $^8$Be, and $^4$He-$^4$He scattering ---
could determine whether a resonant or virtual shallow state at LO 
is transformed into a bound state by subleading interactions,
thus elucidating the relation between clusterization and QCD.

In order to better appreciate the cluster nature of 
our solution for $^{16}$O, 
we have studied the radial nucleon density and the sampled probability density
for the nucleons. In both cases the occurrence of clusterization is evident. 
From our results it is not possible to infer any significant correlation 
between the clusters, which once more confirms the extremely weak interaction 
among them within 
\eftnopi. 
We would like to point out that localization was not imposed in the wave 
function used to project out the ground state; rather, it spontaneously arises 
from the optimization procedure (despite the correlations 
being fully translationally invariant) and it is preserved by the subsequent 
imaginary-time projection. 

Current QMC (AFDMC) results have now reached an accuracy
level that allows for discussing the few-MeV energies
involved in this class of phenomena, which are relevant
for a deeper understanding of how the systematics in nuclear
physics arises from QCD.
Starting from LCQD data obtained for values of $m_\pi$ smaller than 
the ones employed in this 
work, and yet larger than the physical one, would allow us to establish the 
threshold for which nuclei as large as $^{16}$O are 
stable against the breakup into four $^4$He 
clusters, if such a threshold exists. To perform this analysis,
it is essential to include 
higher-order terms in the 
\eftnopi~interaction, possibly up to N$^2$LO, where tensor contributions 
appear. This also  
requires a substantial improvement of the existing LQCD data on light nuclei, 
which, even for large $m_\pi$, are currently affected by statistical errors 
that do not allow for an effective constraint of the interaction parameters.

\section*{Acknowledgements}
We would like to thank N. Barnea, D. Gazit, G. Orlandini, 
and W. Leidemann
for useful discussions about the subject of this paper.  
This research was conducted in the scope of the Laboratoire international 
associ\'e (LIA) COLL-AGAIN and supported 
in part
by the U.S. Department of Energy, Office of Science, Office of Nuclear Physics,
under contracts DE-AC02-06CH11357 (A.L.)
and DE-FG02-04ER41338 (U.v.K.),
and by
the European Union Research and Innovation program Horizon 2020
under grant No. 654002 (U.v.K.).
The work of A.R. was supported by NSF Grant No.\ AST-1333607.
J.K. acknowledges support by the NSF Grant No. PHY15-15738.
Under an award of computer time provided by the INCITE program, 
this research used resources of the Argonne Leadership Computing Facility 
at Argonne National Laboratory, which is supported by the Office of Science 
of the U.S. Department of Energy under contract DE-AC02-06CH11357. 

\section*{Appendix: Statistical and systematic error estimation}

The procedure we adopted in order to estimate the error in the extrapolations 
performed in this work is as follows. We can distinguish between two sources of
errors. The first is a systematic error corresponding to the choice of neglecting 
the next (cubic) order in the  expansion Eq.~\eqref{eq:cut_off_extrapolation} 
and of removing the initial data point at $\Lambda=2$ fm$^{-1}$.
The second is a statistical error coming from the uncertainties in the data 
used for the extrapolation.

The first kind of error is estimated by considering the maximum spread in three different
extrapolations: two quadratic extrapolations obtained by either neglecting the 
results at $\Lambda=2$ fm$^{-1}$ or by using all available data (the latter is 
included only if the reduced $\chi^2$ is 
$\approx1$) and a cubic extrapolation that uses all data.

For the second 
type of error, it is convenient to write Eq. \eqref{eq:cut_off_extrapolation} 
as a simple quadratic form,
\begin{equation}
\mathcal{O}_\Lambda = O  + \mathcal{C}_0 \Gamma+ \mathcal{C}_1\Gamma^2 + \cdots
\end{equation}
where $\Gamma=1/\Lambda$. Given 
that we have only three pairs 
$(\Lambda,\mathcal{O}_\Lambda)$,
it is straightforward to see that
\begin{equation}
\mathcal{C}_1 = \frac{1}{\Gamma_2-\Gamma_3}
\left[\frac{\mathcal{O}_2-\mathcal{O}_1}{\Gamma_2-\Gamma_1}
-\frac{\mathcal{O}_3-\mathcal{O}_1}{\Gamma_3-\Gamma_1}\right]
\end{equation}
together with
\begin{equation}
\mathcal{C}_0 =\frac{\mathcal{O}_3-\mathcal{O}_1}{\Gamma_3-\Gamma_1} 
- \left(\Gamma_3+\Gamma_1\right)\mathcal{C}_1
\end{equation}
and
\begin{equation}
O =\mathcal{O}_1 - \mathcal{C}_0 \Gamma_1 - \mathcal{C}_1 \Gamma_1^2 \; .
\end{equation}
At this point 
it is simple to estimate the errors by propagation of the measurement 
uncertainty. 
We have
\begin{equation}
\delta \mathcal{C}_1 = \frac{1}{\Gamma_2-\Gamma_3}
\sqrt{\left[\frac{\delta\mathcal{O}_2^2+\delta\mathcal{O}_1^2}
{\left(\Gamma_2-\Gamma_1\right)^2}
+\frac{\delta\mathcal{O}_3^2+\delta\mathcal{O}_1^2}
{\left(\Gamma_3-\Gamma_1\right)^2}\right]}
\end{equation}
and 
\begin{equation}
\delta \mathcal{C}_0 = 
\sqrt{\frac{\delta\mathcal{O}_3^2+\delta\mathcal{O}_1^2}
{\left(\Gamma_3-\Gamma_1\right)^2}
+\left(\Gamma_3+\Gamma_1\right)^2\delta\mathcal{C}_1^2}\, ,
\end{equation}
and then finally
\begin{equation}
\delta O=\sqrt{\delta\mathcal{O}_1^2 
+ \delta \mathcal{C}_0^2 \Gamma_1^2 + \delta \mathcal{C}_1^2 \Gamma_1^4 } \; .
\end{equation}

Both error estimates appear in the results reported in the main text.

\bibliographystyle{elsarticle-num}
\bibliography{biblio}

\begin{thebibliography}{10}
\expandafter\ifx\csname url\endcsname\relax
  \def\url#1{\texttt{#1}}\fi
\expandafter\ifx\csname urlprefix\endcsname\relax\def\urlprefix{URL }\fi
\expandafter\ifx\csname href\endcsname\relax
  \def\href#1#2{#2} \def\path#1{#1}\fi

\bibitem{Beane:2012vq}
S.~R. Beane, E.~Chang, S.~D. Cohen, W.~Detmold, H.~W. Lin, T.~C. Luu,
  K.~Orginos, A.~Parre\~no, M.~J. Savage, A.~Walker-Loud, {Light Nuclei and
  Hypernuclei from Quantum Chromodynamics in the Limit of SU(3) Flavor
  Symmetry}, Phys. Rev. D87~(3) (2013) 034506.
\newblock \href {http://arxiv.org/abs/1206.5219} {\path{arXiv:1206.5219}},
  \href {http://dx.doi.org/10.1103/PhysRevD.87.034506}
  {\path{doi:10.1103/PhysRevD.87.034506}}.

\bibitem{Yamazaki:2012hi}
T.~Yamazaki, K.-i. Ishikawa, Y.~Kuramashi, A.~Ukawa, {Helium nuclei, deuteron
  and dineutron in 2+1 flavor lattice QCD}, Phys. Rev. D86 (2012) 074514.
\newblock \href {http://arxiv.org/abs/1207.4277} {\path{arXiv:1207.4277}},
  \href {http://dx.doi.org/10.1103/PhysRevD.86.074514}
  {\path{doi:10.1103/PhysRevD.86.074514}}.

\bibitem{Bedaque:2002mn}
P.~F. Bedaque, U.~van Kolck, {Effective field theory for few nucleon systems},
  Ann. Rev. Nucl. Part. Sci. 52 (2002) 339--396.
\newblock \href {http://arxiv.org/abs/nucl-th/0203055}
  {\path{arXiv:nucl-th/0203055}}, \href
  {http://dx.doi.org/10.1146/annurev.nucl.52.050102.090637}
  {\path{doi:10.1146/annurev.nucl.52.050102.090637}}.

\bibitem{Epelbaum:2008ga}
E.~Epelbaum, H.-W. Hammer, U.-G. Mei{\ss}ner, {Modern Theory of Nuclear
  Forces}, Rev. Mod. Phys. 81 (2009) 1773--1825.
\newblock \href {http://arxiv.org/abs/0811.1338} {\path{arXiv:0811.1338}},
  \href {http://dx.doi.org/10.1103/RevModPhys.81.1773}
  {\path{doi:10.1103/RevModPhys.81.1773}}.

\bibitem{Machleidt:2011zz}
R.~Machleidt, D.~R. Entem, {Chiral effective field theory and nuclear forces},
  Phys. Rept. 503 (2011) 1--75.
\newblock \href {http://arxiv.org/abs/1105.2919} {\path{arXiv:1105.2919}},
  \href {http://dx.doi.org/10.1016/j.physrep.2011.02.001}
  {\path{doi:10.1016/j.physrep.2011.02.001}}.

\bibitem{Schmidt:1999lik}
K.~E. Schmidt, S.~Fantoni, {A quantum Monte Carlo method for nucleon systems},
  Phys. Lett. B446 (1999) 99--103.
\newblock \href {http://dx.doi.org/10.1016/S0370-2693(98)01522-6}
  {\path{doi:10.1016/S0370-2693(98)01522-6}}.

\bibitem{Barnea:2013uqa}
N.~Barnea, L.~Contessi, D.~Gazit, F.~Pederiva, U.~van Kolck, {Effective Field
  Theory for Lattice Nuclei}, Phys. Rev. Lett. 114~(5) (2015) 052501.
\newblock \href {http://arxiv.org/abs/1311.4966} {\path{arXiv:1311.4966}},
  \href {http://dx.doi.org/10.1103/PhysRevLett.114.052501}
  {\path{doi:10.1103/PhysRevLett.114.052501}}.

\bibitem{Beane:2015yha}
S.~R. Beane, E.~Chang, W.~Detmold, K.~Orginos, A.~Parre\~no, M.~J. Savage,
  B.~C. Tiburzi, {Ab initio Calculation of the np→dγ Radiative Capture
  Process}, Phys. Rev. Lett. 115~(13) (2015) 132001.
\newblock \href {http://arxiv.org/abs/1505.02422} {\path{arXiv:1505.02422}},
  \href {http://dx.doi.org/10.1103/PhysRevLett.115.132001}
  {\path{doi:10.1103/PhysRevLett.115.132001}}.

\bibitem{Kirscher:2015yda}
J.~Kirscher, N.~Barnea, D.~Gazit, F.~Pederiva, U.~van Kolck, {Spectra and
  Scattering of Light Lattice Nuclei from Effective Field Theory}, Phys. Rev.
  C92~(5) (2015) 054002.
\newblock \href {http://arxiv.org/abs/1506.09048} {\path{arXiv:1506.09048}},
  \href {http://dx.doi.org/10.1103/PhysRevC.92.054002}
  {\path{doi:10.1103/PhysRevC.92.054002}}.

\bibitem{Chen:1999tn}
J.-W. Chen, G.~Rupak, M.~J. Savage, {Nucleon-nucleon effective field theory
  without pions}, Nucl. Phys. A653 (1999) 386--412.
\newblock \href {http://arxiv.org/abs/nucl-th/9902056}
  {\path{arXiv:nucl-th/9902056}}, \href
  {http://dx.doi.org/10.1016/S0375-9474(99)00298-5}
  {\path{doi:10.1016/S0375-9474(99)00298-5}}.

\bibitem{Kong:1999sf}
X.~Kong, F.~Ravndal, {Coulomb effects in low-energy proton proton scattering},
  Nucl. Phys. A665 (2000) 137--163.
\newblock \href {http://arxiv.org/abs/hep-ph/9903523}
  {\path{arXiv:hep-ph/9903523}}, \href
  {http://dx.doi.org/10.1016/S0375-9474(99)00406-6}
  {\path{doi:10.1016/S0375-9474(99)00406-6}}.

\bibitem{Vanasse:2013sda}
J.~Vanasse, {Fully Perturbative Calculation of $nd$ Scattering to
  Next-to-next-to-leading-order}, Phys. Rev. C88~(4) (2013) 044001.
\newblock \href {http://arxiv.org/abs/1305.0283} {\path{arXiv:1305.0283}},
  \href {http://dx.doi.org/10.1103/PhysRevC.88.044001}
  {\path{doi:10.1103/PhysRevC.88.044001}}.

\bibitem{Konig:2015aka}
S.~K{\"o}nig, H.~W. Grie{\ss}hammer, H.-W. Hammer, U.~van Kolck, {Effective
  theory of $^3$H and $^3$He}, J. Phys. G43~(5) (2016) 055106.
\newblock \href {http://arxiv.org/abs/1508.05085} {\path{arXiv:1508.05085}},
  \href {http://dx.doi.org/10.1088/0954-3899/43/5/055106}
  {\path{doi:10.1088/0954-3899/43/5/055106}}.

\bibitem{Platter:2004zs}
L.~Platter, H.-W. Hammer, U.-G. Mei{\ss}ner, {On the correlation between the
  binding energies of the triton and the alpha-particle}, Phys. Lett. B607
  (2005) 254--258.
\newblock \href {http://arxiv.org/abs/nucl-th/0409040}
  {\path{arXiv:nucl-th/0409040}}, \href
  {http://dx.doi.org/10.1016/j.physletb.2004.12.068}
  {\path{doi:10.1016/j.physletb.2004.12.068}}.

\bibitem{Stetcu:2006ey}
I.~Stetcu, B.~R. Barrett, U.~van Kolck, {No-core shell model in an
  effective-field-theory framework}, Phys. Lett. B653 (2007) 358--362.
\newblock \href {http://arxiv.org/abs/nucl-th/0609023}
  {\path{arXiv:nucl-th/0609023}}, \href
  {http://dx.doi.org/10.1016/j.physletb.2007.07.065}
  {\path{doi:10.1016/j.physletb.2007.07.065}}.

\bibitem{Bazak:2016wxm}
B.~Bazak, M.~Eliyahu, U.~van Kolck, {Effective Field Theory for Few-Boson
  Systems}, Phys. Rev. A94~(5) (2016) 052502.
\newblock \href {http://arxiv.org/abs/1607.01509} {\path{arXiv:1607.01509}},
  \href {http://dx.doi.org/10.1103/PhysRevA.94.052502}
  {\path{doi:10.1103/PhysRevA.94.052502}}.

\bibitem{Bedaque:1997qi}
P.~F. Bedaque, U.~van Kolck, {Nucleon deuteron scattering from an effective
  field theory}, Phys. Lett. B428 (1998) 221--226.
\newblock \href {http://arxiv.org/abs/nucl-th/9710073}
  {\path{arXiv:nucl-th/9710073}}, \href
  {http://dx.doi.org/10.1016/S0370-2693(98)00430-4}
  {\path{doi:10.1016/S0370-2693(98)00430-4}}.

\bibitem{Bedaque:1999ve}
P.~F. Bedaque, H.-W. Hammer, U.~van Kolck, {Effective theory of the triton},
  Nucl. Phys. A676 (2000) 357--370.
\newblock \href {http://arxiv.org/abs/nucl-th/9906032}
  {\path{arXiv:nucl-th/9906032}}, \href
  {http://dx.doi.org/10.1016/S0375-9474(00)00205-0}
  {\path{doi:10.1016/S0375-9474(00)00205-0}}.

\bibitem{vanKolck:1998}
U.~van Kolck, {Effective field theory of short range forces}, Nucl. Phys. A645
  (1999) 273--302.
\newblock \href {http://arxiv.org/abs/nucl-th/9808007}
  {\path{arXiv:nucl-th/9808007}}, \href
  {http://dx.doi.org/10.1016/S0375-9474(98)00612-5}
  {\path{doi:10.1016/S0375-9474(98)00612-5}}.

\bibitem{Bedaque:1998kg}
P.~F. Bedaque, H.-W. Hammer, U.~van Kolck, {Renormalization of the three-body
  system with short range interactions}, Phys. Rev. Lett. 82 (1999) 463--467.
\newblock \href {http://arxiv.org/abs/nucl-th/9809025}
  {\path{arXiv:nucl-th/9809025}}, \href
  {http://dx.doi.org/10.1103/PhysRevLett.82.463}
  {\path{doi:10.1103/PhysRevLett.82.463}}.

\bibitem{Binder:2015mbz}
S.~Binder, et~al., {Few-nucleon systems with state-of-the-art chiral
  nucleon-nucleon forces}, Phys. Rev. C93~(4) (2016) 044002.
\newblock \href {http://arxiv.org/abs/1505.07218} {\path{arXiv:1505.07218}},
  \href {http://dx.doi.org/10.1103/PhysRevC.93.044002}
  {\path{doi:10.1103/PhysRevC.93.044002}}.

\bibitem{Carlson:2014vla}
J.~Carlson, S.~Gandolfi, F.~Pederiva, S.~C. Pieper, R.~Schiavilla, K.~E.
  Schmidt, R.~B. Wiringa, {Quantum Monte Carlo methods for nuclear physics},
  Rev. Mod. Phys. 87 (2015) 1067.
\newblock \href {http://arxiv.org/abs/1412.3081} {\path{arXiv:1412.3081}},
  \href {http://dx.doi.org/10.1103/RevModPhys.87.1067}
  {\path{doi:10.1103/RevModPhys.87.1067}}.

\bibitem{Gandolfi:2014ewa}
S.~Gandolfi, A.~Lovato, J.~Carlson, K.~E. Schmidt, {From the lightest nuclei to
  the equation of state of asymmetric nuclear matter with realistic nuclear
  interactions}, Phys. Rev. C90~(6) (2014) 061306.
\newblock \href {http://arxiv.org/abs/1406.3388} {\path{arXiv:1406.3388}},
  \href {http://dx.doi.org/10.1103/PhysRevC.90.061306}
  {\path{doi:10.1103/PhysRevC.90.061306}}.

\bibitem{Sorella:2005}
S.~Sorella, \href{http://link.aps.org/doi/10.1103/PhysRevB.71.241103}{{Wave
  function optimization in the variational Monte Carlo method}}, Phys. Rev. B
  71 (2005) 241103.
\newblock \href {http://dx.doi.org/10.1103/PhysRevB.71.241103}
  {\path{doi:10.1103/PhysRevB.71.241103}}.
\newline\urlprefix\url{http://link.aps.org/doi/10.1103/PhysRevB.71.241103}

\bibitem{Lagaris:1981mn}
I.~E. Lagaris, V.~R. Pandharipande, {Variational Calculations of Realistic
  Models of Nuclear Matter}, Nucl. Phys. A359 (1981) 349--364.
\newblock \href {http://dx.doi.org/10.1016/0375-9474(81)90241-4}
  {\path{doi:10.1016/0375-9474(81)90241-4}}.

\bibitem{Arriaga:1995ue}
A.~Arriaga, V.~R. Pandharipande, R.~B. Wiringa, {Three body correlations in few
  body nuclei}, Phys. Rev. C52 (1995) 2362--2368.
\newblock \href {http://arxiv.org/abs/nucl-th/9506036}
  {\path{arXiv:nucl-th/9506036}}, \href
  {http://dx.doi.org/10.1103/PhysRevC.52.2362}
  {\path{doi:10.1103/PhysRevC.52.2362}}.

\bibitem{Toulouse:2007}
J.~{Toulouse}, C.~J. {Umrigar}, {Optimization of quantum Monte Carlo wave
  functions by energy minimization}, J. Chem. Phys. 126~(8) (2007) 084102.
\newblock \href {http://arxiv.org/abs/physics/0701039}
  {\path{arXiv:physics/0701039}}, \href {http://dx.doi.org/10.1063/1.2437215}
  {\path{doi:10.1063/1.2437215}}.

\bibitem{Umrigar:2005}
C.~J. Umrigar, C.~Filippi, Energy and variance optimization of many-body wave
  functions, Phys. Rev. Lett. 94 (2005) 150201.
\newblock \href {http://dx.doi.org/10.1103/PhysRevLett.94.150201}
  {\path{doi:10.1103/PhysRevLett.94.150201}}.

\bibitem{Lee:2005}
M.~W. Lee, M.~Mella, A.~M. Rappe, {Electronic quantum Monte Carlo calculations
  of atomic forces, vibrations, and anharmonicities}, J. Chem. Phys. 122~(24)
  (2005) 244103.
\newblock \href {http://dx.doi.org/10.1063/1.1924690}
  {\path{doi:10.1063/1.1924690}}.

\bibitem{Ottermann:1985km}
C.~R. Ottermann, G.~Kobschall, K.~Maurer, K.~Rohrich, C.~Schmitt, V.~H.
  Walther, {ELASTIC ELECTRON SCATTERING FROM HE-3 AND HE-4}, Nucl. Phys. A436
  (1985) 688--698.
\newblock \href {http://dx.doi.org/10.1016/0375-9474(85)90554-8}
  {\path{doi:10.1016/0375-9474(85)90554-8}}.

\bibitem{Lynn:2014zia}
J.~E. Lynn, J.~Carlson, E.~Epelbaum, S.~Gandolfi, A.~Gezerlis, A.~Schwenk,
  {Quantum Monte Carlo Calculations of Light Nuclei Using Chiral Potentials},
  Phys. Rev. Lett. 113~(19) (2014) 192501.
\newblock \href {http://arxiv.org/abs/1406.2787} {\path{arXiv:1406.2787}},
  \href {http://dx.doi.org/10.1103/PhysRevLett.113.192501}
  {\path{doi:10.1103/PhysRevLett.113.192501}}.

\bibitem{Wiringa:2000gb}
R.~B. Wiringa, S.~C. Pieper, J.~Carlson, V.~R. Pandharipande, {Quantum Monte
  Carlo calculations of A = 8 nuclei}, Phys. Rev. C62 (2000) 014001.
\newblock \href {http://arxiv.org/abs/nucl-th/0002022}
  {\path{arXiv:nucl-th/0002022}}, \href
  {http://dx.doi.org/10.1103/PhysRevC.62.014001}
  {\path{doi:10.1103/PhysRevC.62.014001}}.

\bibitem{McIlroy:2017ssf}
C.~McIlroy, C.~Barbieri, T.~Inoue, T.~Doi, T.~Hatsuda, {Doubly magic nuclei
  from Lattice QCD forces at $M_{PS}=$469 MeV/c$^2$}\href
  {http://arxiv.org/abs/1701.02607} {\path{arXiv:1701.02607}}.

\end{thebibliography}

\end{document}